\documentclass[graybox]{svmult}

\usepackage{mathptmx}       
\usepackage{helvet}         
\usepackage{courier}        
\usepackage{type1cm}        
\usepackage{makeidx}         
\usepackage{graphicx}        
\usepackage{multicol}        
\usepackage[bottom]{footmisc}

\makeindex             
\begin{document}

\title*{The Future of Ultracool Dwarf Science with JWST}
\author{Mark S. Marley and S.K. Leggett}
\institute{Mark Marley \at NASA Ames Research Center, Mail Stop 245-3, Moffett Field CA 94035, \email{Mark.S.Marley@NASA.gov}
\and S. Leggett \at Gemini Observatory, 670 North A'ohoku Place Hilo, HI 96720 \email{skl@gemini.edu}}
%
%
\maketitle

\abstract*{Each chapter should be preceded by an abstract (10--15 lines long) that summarizes the content. The abstract will appear \textit{online} at \url{www.SpringerLink.com} and be available with unrestricted access. This allows unregistered users to read the abstract as a teaser for the complete chapter. As a general rule the abstracts will not appear in the printed version of your book unless it is the style of your particular book or that of the series to which your book belongs.
Please use the 'starred' version of the new Springer \texttt{abstract} command for typesetting the text of the online abstracts (cf. source file of this chapter template \texttt{abstract}) and include them with the source files of your manuscript. Use the plain \texttt{abstract} command if the abstract is also to appear in the printed version of the book.}

\abstract{Ultracool dwarfs exhibit a remarkably varied set of characteristics which hint at the complex physical processes acting in their atmospheres and interiors.   Spectra of these objects not only depend upon their mass and effective temperature, but also their atmospheric chemistry, weather, and dynamics.  As a consequence divining their mass, metallicity and age solely from their spectra has been a challenge.  JWST, by illuminating spectral blind spots and observing objects with constrained masses and ages should finally unearth a sufficient  number of ultracool dwarf Rosetta Stones to allow us to decipher the processes underlying the complex brown dwarf cooling sequence.  In addition the spectra of objects invisible from the ground, including very low mass objects in clusters and nearby cold dwarfs from the disk population, will be seen for the first time.  In combination with other ground- and
space-based assets and programs, JWST  will usher in a new golden era of brown dwarf science and discovery.}

\section{Introduction}

The explosive growth of brown dwarf and ultracool dwarf discoveries over the past dozen years has been so extraordinary that it is a rare paper in the field that does not open by remarking upon it.  The first undisputed brown dwarf, Gl 229 B, was discovered as a companion to an M dwarf in 1995 (Nakajima et al. 1995).
The Two-Micron All Sky Survey (2MASS, Skrutskie et al. 2006)) and the 
Sloan Digital Sky Survey (SDSS, York et al. 2000) 
subsequently revealed large numbers of ultracool low-mass field
dwarfs.  These surveys first led to the discovery of the isolated field 
late-L dwarfs (Kirkpatrick et al. 1999), then the mid-T dwarfs (Burgasser et al. 1999,
Strauss et al. 1999) and finally the early-T dwarfs (Leggett et al. 2000).
Today over 600 warm ($T_{\rm eff} \sim 2400$ to 1400 K) L and cool (600 to 1400 K) T dwarfs are known\footnote{See http://www.DwarfArchives.org} and the quest for the elusive, even cooler ``Y'' dwarfs is ongoing.  

Note that collectively late M and later type dwarfs are often termed `Ultracool Dwarfs' (or UCDs) to avoid having to distinguish whether particular warm objects in this group are above or below the hydrogen burning minimum mass, the  requirement for bestowing the term `brown dwarf'.  Brown dwarfs will continuously cool over time.  The more massive UCDs will eventually arrive on the bottom of the hydrogen burning main sequence (Burrows et al. 1997).

\begin{figure}[ht]
\centering
\includegraphics[scale=.85,angle=0]{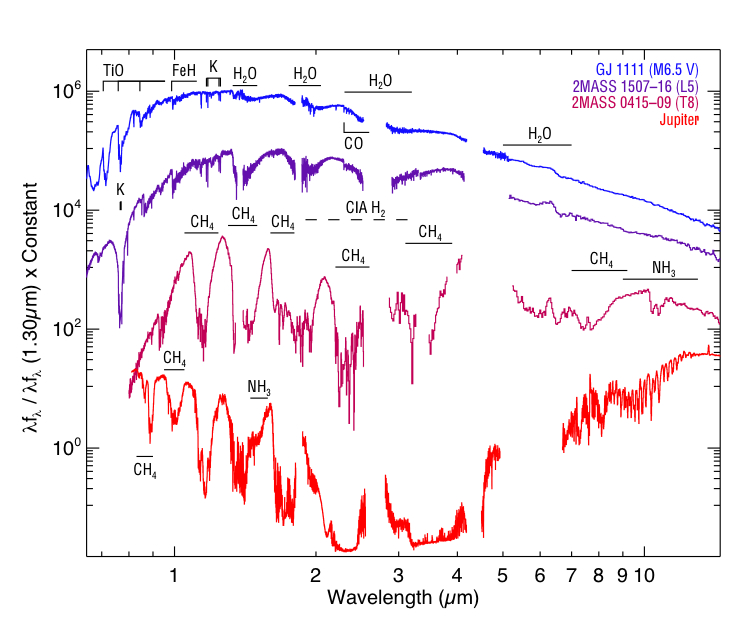}
\caption{The most prominent signatures of the ultra cool dwarf spectral sequence are seen in these 0.65 to $14.5\,\rm \mu m$ spectra of a mid-M, L, and T dwarfs as well as Jupiter (adapted from Cushing et al. (2006)).  The spectra have been normalized to unity at $1.3\,\rm \mu m$ and multiplied by constants.  Major absorption bands are marked. The collision-induced opacity of $\rm H_2$ is indicated as a dashed line because it shows no distinct spectral features but rather a broad, smooth absorption.  Jupiter's flux shortward of $\sim 4\,\rm \mu m$ is predominantly scattered solar light; thermal emission dominates at longer wavelengths (near- and mid-infrared Jovian spectra from Rayner, Cushing \& Vacca (in preparation) and Kunde et al. (2004), respectively).}
\label{fig:argh}
\end{figure}

Ultracool dwarf science is exciting not only for the rapid pace of discovery, but for a host of other reasons as well.  First, since brown dwarfs lack an internal energy source (beyond a brief period of deuterium burning), they cool off over time; they thus reach effective temperatures below those found in stars and enter the realm where chemical equilibrium  favors such decidedly `unstellar' atmospheric species as $\rm CH_4$ and $\rm NH_3$.  Along with these more typical `planetary' gasses, silicate and iron clouds are found in their atmospheres, leading to interesting, complex interactions between atmospheric chemical, radiative-transfer, dynamical, and meteorological processes.   Ultracool dwarfs thus bridge the domain between the bottom of the stellar main sequence and giant planets.  They are a laboratory for understanding processes that will also be important in the characterization of extrasolar giant planets.  Second, they occupy the low mass end of the stellar initial mass function. Understanding the IMF requires that we understand the masses of individual field objects, which ultimately requires an understanding of their luminosity evolution as well as the dependence of their spectra on  mass, gravity, effective temperature and metallicity.  Finally, as {\it terra incognita}, brown dwarfs (some of our nearest stellar neighbors) have offered a series of surprises that test our ability to understand the universe around us.

\begin{figure}[ht]
\centering
\includegraphics[scale=.80,angle=0]{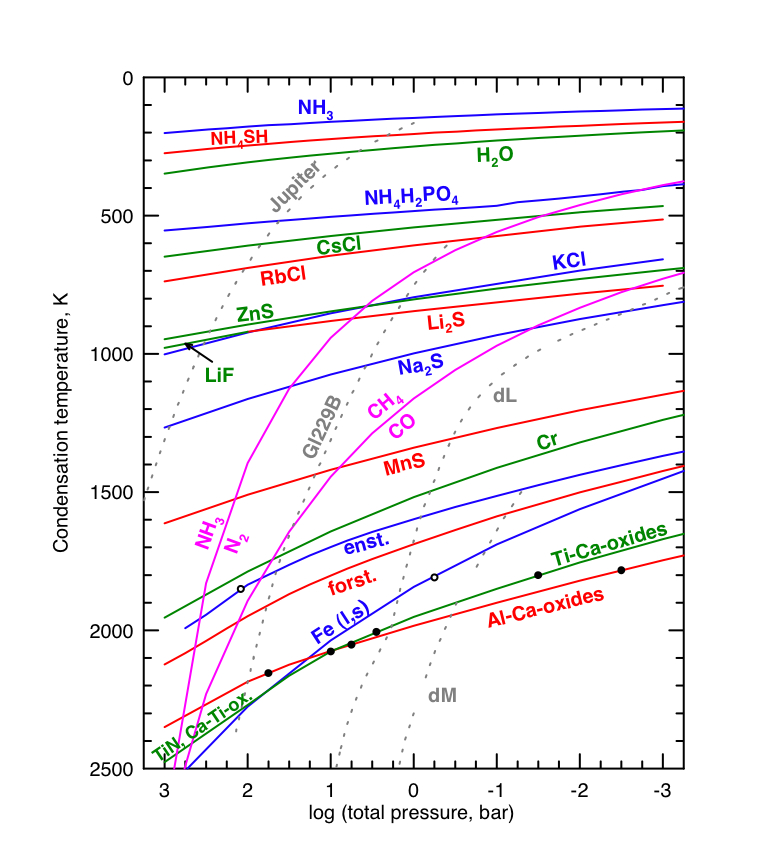}
\caption{Important chemical equilibrium boundaries for substellar objects (modified from Lodders \& Fegley 2006).  Green, red, and blue lines denote various condensation boundaries for a solar abundance mixture of gasses in a substellar atmosphere.   Light purple lines denote equilibrium boundaries between important gaseous species.  Grey dashed lines show model atmospheric temperature-pressure profiles for  M, L and T dwarfs (the latter specifically for Gl 229 B) as well as for Jupiter. As one moves upwards in the diagram along the model $(T,P)$ curves, the labeled species will condense at the intersection with the condensation curves and would be expected to be absent from the gas at lower temperatures further up along the model profiles.  This figure can be compared with Figure \ref{fig:argh} to understand why spectral features for various compounds are present or absent in each observed spectrum.}
\label{fig:chem}
\end{figure}

The most distinctive features of the UCD spectral sequence are highlighted in 
Figure~\ref{fig:argh}. At effective temperatures  below
those of late-M dwarfs, several chemical changes (illustrated in Figure \ref{fig:chem}) occur that strongly impact the spectral
energy distribution.  First, major diatomic metal species (particularly TiO and FeH) become incorporated into grains, leading to the gradual departure of hallmarks of the M spectral sequence (Kirkpatrick et al. 1999).  Second, the formation of iron and silicate grains produces optically thick clouds 
 that veil gaseous absorption bands and redden the
near-IR $JHK$ colors of L dwarfs.  The atmospheric temperature domain where these clouds are most important is 
$\sim 1500-2000\, \rm K$ (e.g. Ackerman \& Marley 2001).  At lower $T_{\rm eff}$, the
clouds lie near or below the base of the wavelength-dependent photosphere, and
only marginally affect the SEDs of T dwarfs.  Finally, CH$_4$
supplants CO as the dominant carbon-bearing molecule.  This transition is first
noted in the 3--4~$\mu$m spectra of mid-L dwarfs (Noll et al.\ 2000) and appears
in both the $H$ and $K$ bands of T0 dwarfs (Geballe et al.\ 2002).  Together,
increasing CH$_4$ absorption and sinking cloud decks cause progressively bluer
near-IR colors of T dwarfs.  For types T5 and later, significant 
collision-induced H$_2$ opacity in the $K$ band enhances the trend toward bluer near-infrared colors.

\begin{figure}[ht]
\centering
\includegraphics[scale=0.3,angle=0]{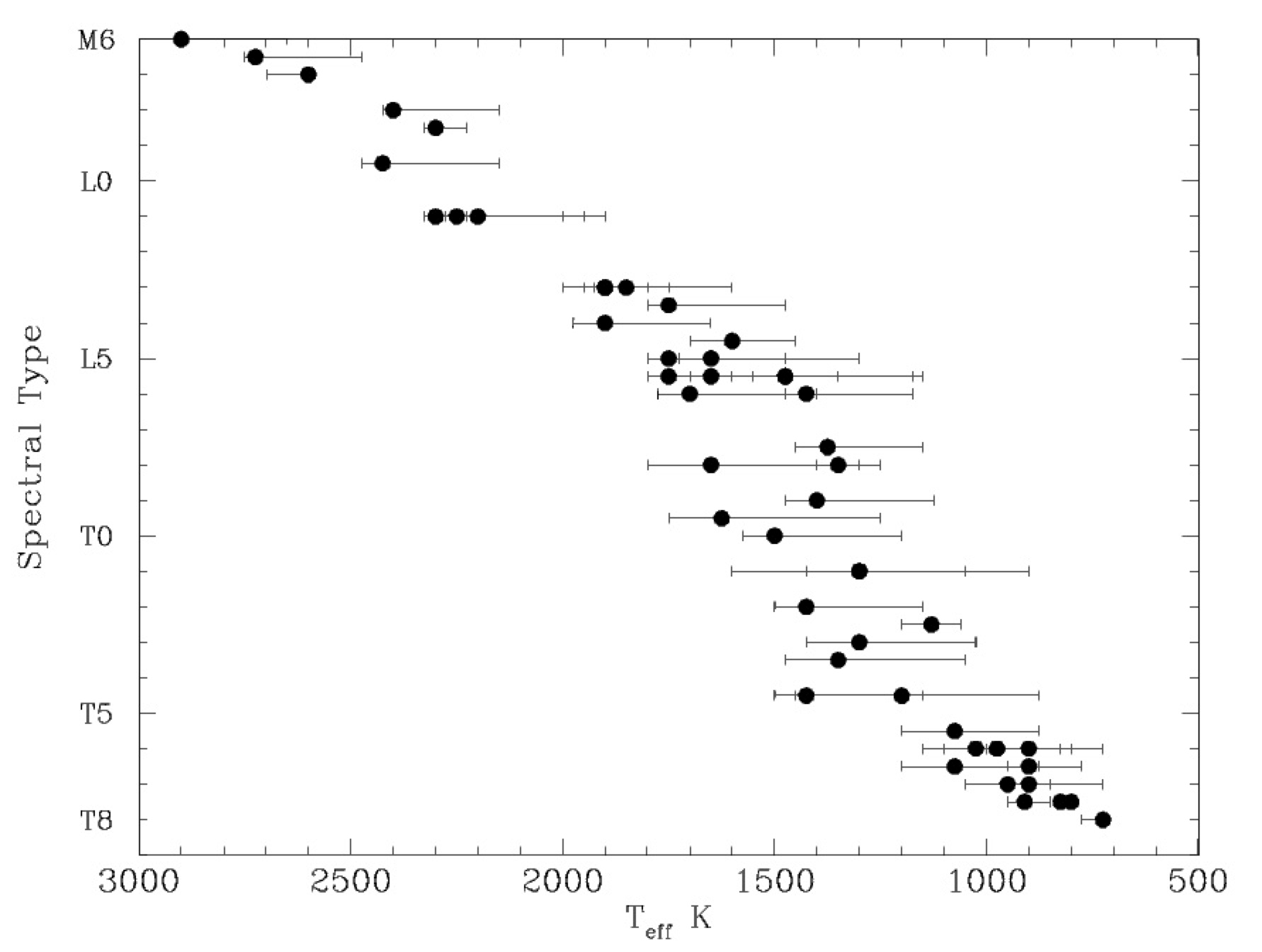}
\caption{Effective temperature as a function of infrared spectral type for ultracool dwarfs with known parallax (data from Golimowski et al.\ 2004, Vrba et al.\ 2004, and Luhman et al.\ 2007).  Note the roughly constant effective temperature for dwarfs of spectral types from late L to early T. See Kirkpatrick (2007) for further discussion.}
\label{fig:teffs}
\end{figure}

These changes in spectral features are used to assign spectral types to UCDs as briefly explained in \S2.  Given assigned spectral types, measurement of the bolometric luminosity of individual objects along with their parallaxes connect the spectral sequence to effective temperature.   
Figure~\ref{fig:teffs} illustrates the effective temperature as a function of spectral type from late M through late T.  While the general correlation of increasing spectral type with falling effective temperature is unmistakable, a remarkably rapid set  of spectral changes (as expressed in the variation in spectral type) happens over a relatively small span of $T_{\rm eff}$ near 1400 K.  As we will discuss, understanding this variation in expressed spectral signatures, the `L to T transition', is a key subject of current brown dwarf research.

Most of the scientific inquiry into these ultracool dwarfs has focused on their formation and youth, on the resultant initial mass function, and on the determination of their global properties, particularly mass, effective temperature, metallicity, and cloudiness. This review will focus primarily on the latter areas. Burrows et al. (2001) provide a much more in depth review and background to brown dwarf science and is an excellent starting point for those new to the subject.  Kirkpatrick (2007) provides a more current look at outstanding issues in the field from an observational perspective.  Here we first present a very brief review of the ultracool dwarf spectral types and the nature of the current datasets.  We then move on to discuss the role that clouds and atmospheric mixing play in controlling the emitted spectra of these objects and the enigmatic L- to T-type transition. Because  clouds control the spectral energy distribution of the L and early T dwarfs, and since clouds are inherently difficult to model, constraining gravity solely by comparison of observations to spectral data is particularly challenging.  Finally we will close with a look forward to some of the ultracool dwarf science opportunities that will be enabled  by JWST.  Because of space limitations we neglect several other important avenues of UCD research, including studies of the IMF, very young objects, and of objects with unusual colors.  

\section{Spectral Type}

The currently known ultracool dwarfs span spectral types from late M, L0 through L9, and T0 through T9.
The TiO and VO bands, which dominate the optical portions of late-M dwarf spectra, 
disappear in the L dwarfs (Kirkpatrick et al. 1999), where metallic oxides are replaced by metallic hydrides and where features due to neutral alkali 
metals are strong.  To systematize such objects Kirkpatrick et al. established spectral indices and defined an optical classification scheme for L 
dwarfs, which is commonly used.  The indices measure the strengths of TiO, VO, CrH, Rb and Cs features as well as a red color 
term, at wavelengths between 0.71 and 0.99 $\mu$m. 

With the discovery of the  T dwarfs, which have very little flux in the optical, Geballe et
al. (2002) defined a near-infrared classification scheme that encompassed both the L and T dwarfs.  The indices measure the 
strength of the H$_2$O and CH$_4$ absorption features between 1.1 and $2\, \mu$m, and for the early L dwarfs a 
red color term is also used which is slightly modified from Kirkpatrick et al. (1999). 

Burgasser et al. (2002a) introduced a near-infrared classification scheme for T dwarfs that was very similar to that of Geballe 
et al.  The two schemes were unified in Burgasser et al. (2006) and this near-infrared scheme is the commonly used scheme for 
typing T dwarfs. For L dwarfs, both the optical Kirkpatrick et al. and the near-infrared Geballe et al. schemes are used; 
these usually produce the same type (the Geballe et al. scheme was pinned to the Kirkpatrick et al. types), but for L dwarfs
with unusual colors they can give significantly different types. Because of this, the classification scheme used for L dwarfs should always be specified.

Leggett et al. (2007, and other work referenced therein) explore the possible signatures of the next spectral type, for which 
the letter Y has been suggested (Kirkpatrick 2005). It is likely that NH$_3$ features will join the familiar water and methane absorption seen in T dwarfs   as the effective temperatures approach 600~K.  In actuality the situation is likely more complex:  it is already known that the atmospheric NH$_3$ abundance (as seen in {\em Spitzer} mid-infrared spectra of  T dwarfs) is reduced by vertical
mixing which drags N$_2$ up from deeper layers in the atmosphere (e.g., Saumon et al. (2006)).  If this mechanism continues to act at
low effective temperatures (which is not a certainty (Hubeny \& Burrows 2007)) the near-infrared  ammonia features may be weaker than 
expected.   Another outstanding 
problem with predicting the signature of the proposed Y dwarfs, is that the linelist for NH$_3$ is very incomplete at 1.0-1.5 
$\mu$m.  Since the near-infrared flux of Y dwarfs is expectedly to rapidly `collapse' with falling $T_{\rm eff}$ (Burrows et al. 2003), it may even be appropriate to ultimately type these objects with mid-, instead of near-infrared, spectra. In this case spectra obtained by {\em Spitzer} or JWST would be required for spectral typing. Regardless, dwarfs with $T_{\rm 
eff}\sim 650\,\rm K$ are now being  found (e.g. Warren et al. 2007), and it is likely that soon  temperatures where significant 
spectral changes occur will be reached.

\section{Ultracool Dwarf Datasets}

The L and T dwarfs were discovered primarily as a result of the far-red and near-infrared Sloan Digital Sky Survey and 2 
Micron All Sky Survey (e.g. Kirkpatrick et al. 1999, Strauss et al. 1999).  This continues with current surveys - the 
Canada France Hawaii Brown Dwarf Survey and the UKIRT Infrared Deep Sky Survey (e.g., Lodieu et al. 2007) are 
identifying extreme-T dwarfs by their very red far-red  and blue near-infrared colors. Spectral classification is carried out 
in the far-red or near-infrared.  Hence the existing data for L and T dwarfs is primarily far-red and near-infrared imaging 
and spectroscopy.  The spectroscopy has been medium- or low-resolution, both because that is all that is required for the 
spectral classification, but also because the dwarfs are faint.

Some ground-based imaging and spectroscopy has been carried out at 3.0-5.0 $\mu$m (e.g. Noll et al. 2000, 
Golimowski et  al. 2004).   Such work is extremely challenging due to the very high and rapidly variable sky background at 
these wavelengths, and only the brightest dwarfs could be observed from the ground.

This situation changed with the launch of the {\em Spitzer Space Telescope}.  Roellig et al. (2004) and Patten et al. 
(2006) demonstrated the quality and quantity of mid-infrared imaging and spectroscopy of L and T dwarfs that {\em Spitzer} could 
produce, and such work continues through the current, final, cryogenic cycle.  IRS spectral data, which span 6 to 15 $\mu$m, show a strong 11 $\mu$m NH$_3$ absorption
feature in T dwarfs, as well as H$_2$O and CH$_4$ absorption features in both L and T dwarfs  (Figure \ref{fig:argh}).   IRAC photometry covers the 3 to 8 
$\mu$m wavelength range, and the 3 to 5 $\mu$m range may continue to be available in the warm-{\em Spitzer} era.
    These IRAC bandpasses include
CH$_4$, CO and H$_2$O features, and signatures of  vertical transport have been recognized in the photometry (Leggett et al.~2007).

Warren et al. (2007) further suggest that the $H -$ [4.49] color may be a very good indicator of temperature for dwarfs cooler than 
1000~K.  For extreme-T dwarfs the near-infrared CH$_4$ and H$_2$O bands are so strong that it will be difficult to measure
an increase in their strength, hence the mid-infrared may prove to be vital to interpreting the cold objects. 

\section{Ultracool Dwarf Atmospheres}
\begin{figure}[ht]
\centering
\includegraphics[scale=0.22,angle=0]{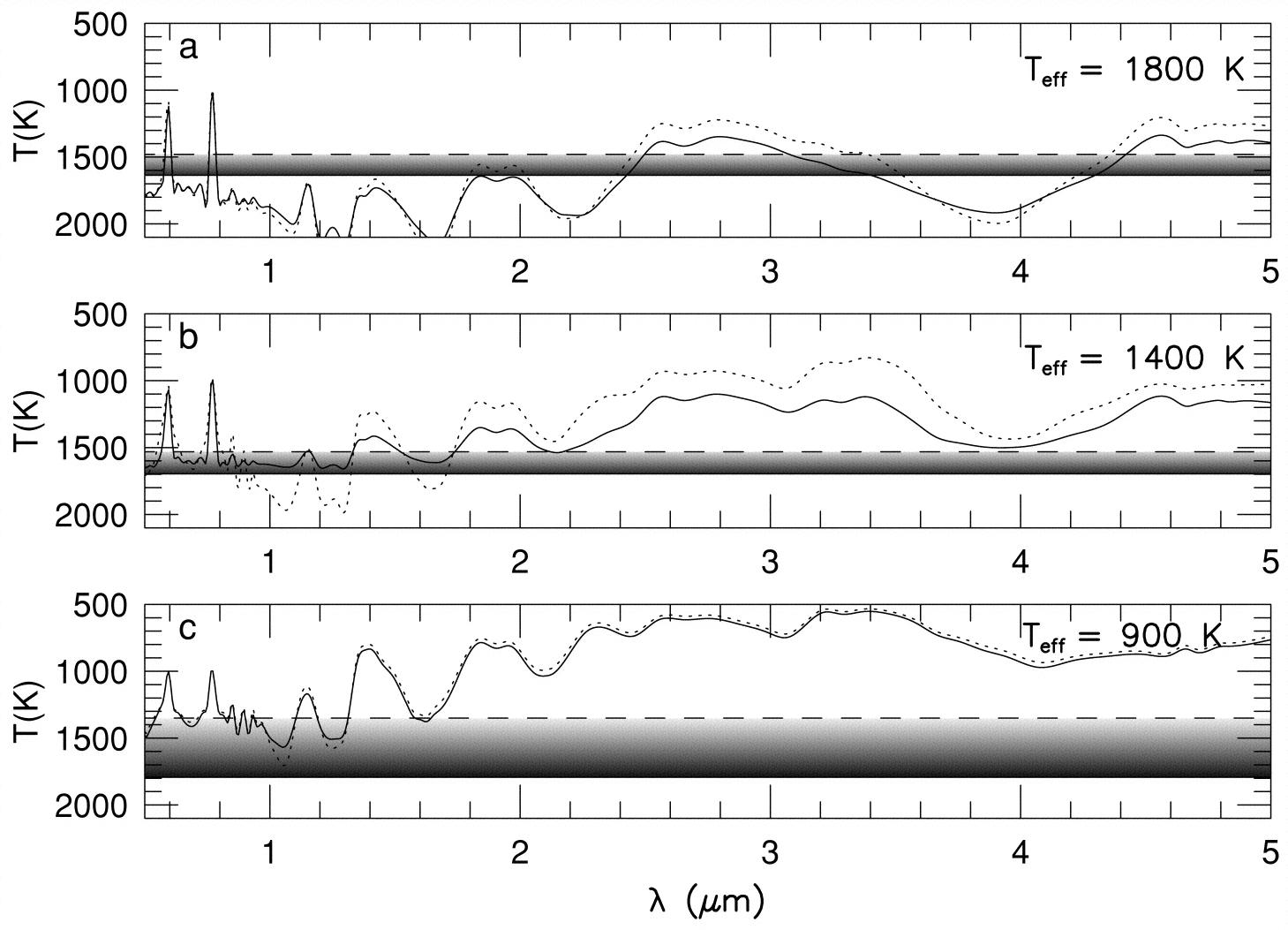}
\caption{Brightness temperature as a function of wavelength for atmosphere models which include (solid) or exclude (dotted) silicate and iron clouds (Ackerman \& Marley 2001).  Brightness temperature increases downward to suggest increasing depth in the atmosphere from which the wavelength-dependent flux emerges. The solid straight line indicates the base of the silicate cloud while the long dashed line denotes the `top' of the cloud (the level in the atmosphere at which the cloud column extinction reaches 0.1). Shading suggests the decrease in cloud extinction with altitude. Since cloud particle radii exceed $10\,\rm \mu m$ in these models, the Mie extinction efficiency is not a strong function of wavelength over the range shown. Shown are models characteristic of (a) an early-type L dwarf with  $T_{\rm eff} = 1800\,\rm K$, (b) a late-type L dwarf with $T_{\rm eff} = 1400\,\rm K$, and (c) a T dwarf with  $T_{\rm eff} = 900\,\rm K$ . All of these models are for solar composition and gravity appropriate for a 30 Jupiter-mass brown dwarf. Note that the spectral region just longward of $1\,\rm \mu m$ is particularly sensitive to the cloud opacity. }
\label{fig:ack}
\end{figure}

Ultracool dwarf emergent spectra are controlled by the variation in abundances of important atomic, molecular, and grain absorbers both with height in the atmosphere at a given age and over time as the objects cool.  The major atomic and molecular absorption features are imprinted on spectra that have no true continuum\footnote{Sharp \& Burrows  (2007) and Freedman et al. (2008) discuss the atmospheric opacity sources in detail.}.  Flux emerges from a many-scale-height-thick range of depths in the atmosphere as a function of wavelength.  For example (Fig. \ref{fig:ack}), in an early L dwarf, brightness temperatures\footnote{Brightness temperature (the temperature that a blackbody that emits radiation of the observed intensity at a given wavelength) is commonly used in planetary atmospheres studies to elucidate the temperature of the emitting level in an atmosphere. }
range from 1000 K in the depths of alkali absorption lines (Burrows et al. 2000) in the far red, to well over 2000 K in the molecular windows in between strong water absorption bands.   By providing a continuum opacity source, clouds can limit the flux emerging in some molecular window regions, but not others.  Furthermore the strength of some molecular absorption features, particularly $\rm CH_4$, CO, and $\rm NH_3$, can depend on the strength of mixing in the atmosphere.  Thus a full description of an ultracool dwarf atmosphere hinges on the dwarf's gravity, effective temperature, cloud properties, and mixing.  In this section we summarize the important unsolved problems related to these atmospheres.

\subsection{Clouds}

At high effective temperatures the column abundance of condensates (see Marley 2000 for a discussion of influences on cloud optical depth) is low and the difference between models computed with and without cloud opacity is slight (Figure \ref{fig:ack}).  At lower temperatures, however, the cloud substantially alters the temperature profile of the atmosphere and provides a continuum opacity source that limits the depth to which the usual molecular windows probe into the atmosphere.  By the effective temperature of the mid-T dwarfs, however, most of the flux emerges from above the cloud level and the clouds are again less important.  For the effective temperature range of the mid to late L dwarfs and the early T dwarfs, however, clouds clearly play a very large role in controlling the vertical structure and emergent spectra of brown dwarfs.

Thus any attempt to model brown dwarf atmospheres must include a treatment of clouds.  However clouds are the leading source of uncertainty in terrestrial atmosphere models and are inherently difficult to model.  Their influence depends on the variation of particle size, abundance, and composition with altitude, which in turn depend on the complex interaction of many microphysical processes (e.g., Ackerman \& Marley 2001, Helling et al. 2006).  Fits of model spectra to observational data are  highly sensitive to the treatment of clouds in the underlying atmosphere model and the approaches taken by various modeling groups vary widely (see the comparison study in Helling et al. 2008a).  For example Cushing et al. (2008) demonstrate reasonably accurate fits of model spectra to near- and mid-IR spectra of a sample of L and T dwarfs, but the precise values of effective temperature and gravity obtained from the fits depend entirely upon the cloud sedimentation efficiency (Ackerman \& Marley 2001) assumed.  Since the models are highly dependent on the cloud description, the derived effective temperature and gravities, while plausible, are nevertheless uncertain.  A similar conclusion was reached by Helling et al. (2008b).  Finding a selection of L dwarfs with known $g$ and $T_{\rm eff}$ that could serve as calibrators of the model spectra would be invaluable.  L dwarf companions to main sequence stars with constrained ages, L dwarf binaries with resolved orbits, and L dwarfs in clusters of known ages are all promising targets for such work.

\subsection{Characterizing Clouds}
The spectral range of the InfraRed Spectrometer (IRS) on {\em Spitzer} includes the $10\, \rm \mu m$ silicate feature which arises from the Si-O stretching vibration in silicate grains. The spectral shape and importance of the silicate feature depends on the particle size and composition of the silicate grains. According to phase-equilibrium arguments, in  brown dwarf atmospheres the first expected silicate condensate is forsterite $\rm Mg_2SiO_4$ (Lodders 2002), at $T\sim 1700\,\rm K$ ($P=1\,\rm bar$). Since Mg and Si have approximately equal abundances in a solar composition atmosphere, the condensation of forsterite leaves substantial silicon, present as SiO, in the gas phase. In equilibrium, at temperatures about 50 to 100 K cooler than the forsterite condensation temperature, the gaseous SiO reacts with the forsterite to form enstatite, $\rm MgSiO_3$ (Lodders 2002). The precise vertical distribution of silicate species depends upon the interplay of the atmospheric dynamics and chemistry and such details have yet to be fully modeled, although efforts to improve the detailed cloud modeling continue (e.g., Cooper et al.\ 2003; Woitke \& Helling 2004; Helling \& Woitke 2006; Helling et al. 2008b).

\begin{figure}[ht]
\centering
\includegraphics[scale=0.5]{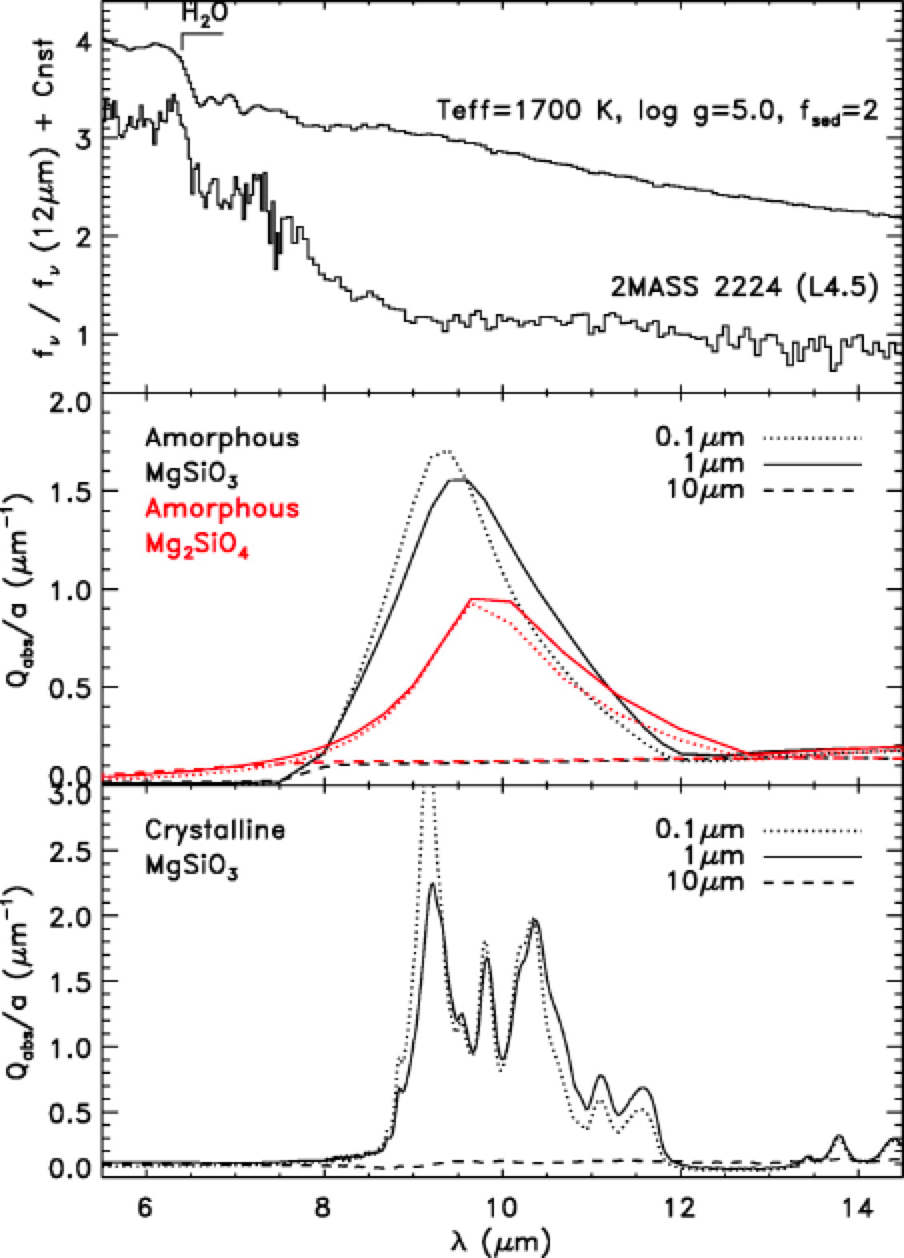}
\caption{ Top: {\em Spitzer} IRS spectrum of 2MASS 2224 (L4.5) and the best fitting model from Cushing et al. (2006). Middle: Optical absorption ($Q_{\rm abs}/a$) for amorphous enstatite ($\rm MgSiO_3$) and forsterite ($\rm Mg_2SiO_3$) for three different particle sizes, 0.1, 1, and $10\,\rm \mu m$. Bottom: Optical absorption for crystalline enstatite, also for three different particle sizes.  The deviation of the model (shifted vertically) from the data suggests that additional small, and perhaps crystalline, silicate grains are required to adequately account for the observed spectrum.}
\label{fig:mie}
\end{figure}

In brown dwarf clouds there is likely a range of particle sizes, ranging from very small, recently condensed grains, to larger grains that have grown by agglomeration. The mean particle size for silicate grains in L dwarf model atmospheres is typically computed to be in the range of several to several tens of microns (Ackerman \& Marley 2001, Helling et al. 2008b) . Figure~\ref{fig:mie} compares the absorption efficiency of silicate grains of various sizes, composition, and crystal structures to the spectrum of 2MASS J2224-0158 (L4.5). For each species the quantity $Q_{\rm abs}/a$, or Mie absorption efficiency divided by particle radius, is shown; all else being equal, the total cloud optical depth is proportional to this quantity (Marley 2000). Large grain sizes tend to have a relatively flat absorption spectra (dashed lines) across the IRS spectral range. Only grains smaller than about $3\,\rm \mu m$ in radius show the classic $10\,\rm \mu m$ silicate feature (Hanner et al. 1994). Figure~\ref{fig:mie} suggests that the mismatch between the models and data may arise from a population of silicate grains that is not captured by the cloud model used to construct the figure (Ackerman \& Marley 2001). The actual silicate cloud may contain both more small particles and a mixture of enstatite and forsterite grains (e.g., Helling \& Woitke 2006), although detailed models for this particular dataset have not been attempted. Furthermore the model shown in the figure employs optical properties of amorphous silicate. It is possible, especially at the higher pressures found in brown dwarf atmospheres, that the grains are crystalline, not amorphous. Indeed laboratory solar-composition condensation experiments produce crystalline, not amorphous, silicates (Toppani et al. 2004). Crystalline grains (Figure \ref{fig:mie}) can have larger and spectrally richer absorption cross sections.

\subsection{The Transition from L to T}
\label{sec:LT}

\begin{figure}[ht]
\centering
\includegraphics[scale=0.4]{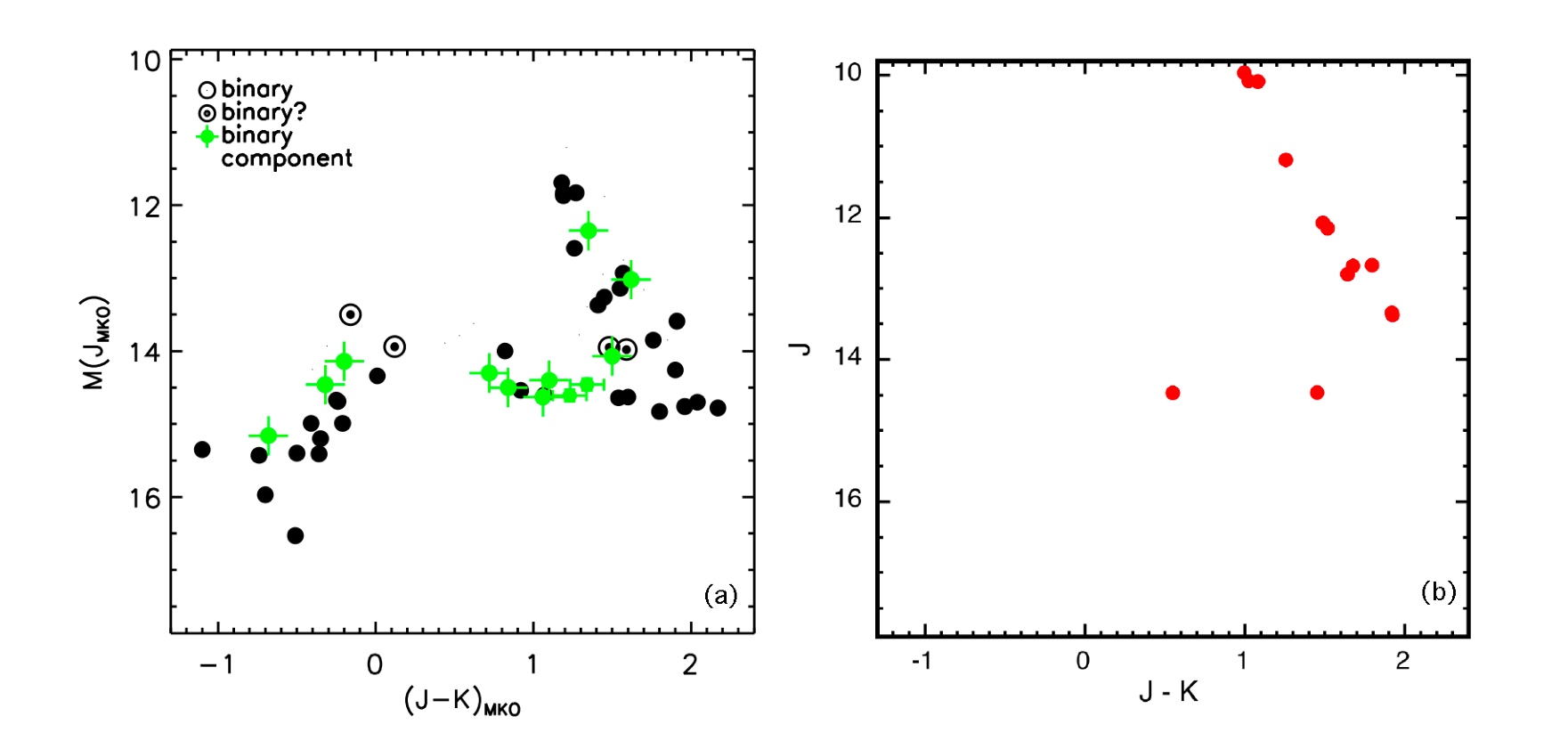}
\caption{Near-infrared color-magnitude diagrams for field and cluster ultracool dwarfs. (a) Black dots show single field L \& T dwarfs.  Green dots are resolved components of binary systems.  Dotted circles are suspected (but unresolved) binaries (figure courtesy M.~Liu based on Liu et al. (2006)). (b) Candidate ultracool dwarfs in the Pleiades in the most sensitive current survey (Casewell et al. 2007).  Faintest objects in this plot have masses of about $11\,\rm M_{Jup}$. Note that at a fixed magnitude the cluster members tend to be redder than the field objects, which is likely a signature of low gravity.  JWST will obtain spectra of quality comparable to Figure \ref{fig:argh} for the candidate objects shown on this panel which will help calibrate evolutionary models of the brown dwarf cooling sequence.  The detection limit for NIRCam on JWST is at about $J=22$ for this cluster  or $\sim 1\,\rm M_{Jup}$ .  Model predictions for colors of objects with $J>15$ are shown in Figure~\ref{fig:y_dwarfs}. }
\label{fig:plds}
\end{figure}

The evolutionary cooling behavior of a given substellar object can be inferred from the field brown dwarf near-infrared color-magnitude diagram (Fig.~\ref{fig:plds}a). Over tens to hundreds of millions of years a given substellar object first moves to redder $J-K$ colors as it cools while falling to fainter J magnitudes.  Around ${\rm M}_J \sim 14 - 15$ the $J-K$ color turns bluer and the $J$ magnitude slightly (and counter-intuitively) brightens (Dahn et al. 2002, Tinney et al. 2003, Vrba et al. 2004).  With further cooling a given dwarf finally falls to fainter $J$ magnitudes and apparently continues to slightly turn somewhat bluer in $J-K$.  The behavior with even greater cooling is as yet uncertain until many more objects with $T_{\rm eff} < 700\,\rm K$ are found.

The `L to T' transition is the `horizontal branch' of the color-magnitude diagram as objects move from red to blue in the diagram.  From Figure~\ref{fig:teffs} we know  that this color (and underlying spectral) change from late-type L dwarfs with $J - K \sim 2.5$ to blue T dwarfs with $J - K \sim -1$  happens rapidly over a small range of effective temperature.  
No brown dwarf evolution model can 
currently reproduce the magnitude of the observed color change over such a small range of $T_{\rm eff}$. Various explanations have 
been suggested including holes forming in the condensate cloud decks (Ackerman \& Marley 2001, Burgasser 
et al. 2002b), an increase in the efficiency of grain sedimentation (Knapp et al. 2004), or a change in particle
size (Burrows et al. 2006).
In a series of papers Tsuji (Tsuji 2002, Tsuji \& Nakajima 2003, Tsuji et al. 2004) proposed that a physically very thin cloud could self-consistently explain the rapid L to T transition.  These models indeed exhibit a somewhat faster L- to T-like transition, but are still not consistent with the observed rapidity of the color change. More recently Tsuji (2005) has favored a  sudden collapse of the global cloud deck at the transition along the lines of the Knapp et al. (2004) suggestion.
 
 Support for a rapid increase in sedimentation efficiency at the L to T transition has 
 come from the model analysis of the 0.8--$14.5\,\mu$m spectra of four transition dwarfs by Cushing et 
al. (2008), who find that the cloud sedimentation efficiency (Ackerman \& Marley 2001) indeed increases across the transition.
The Cushing et al. (2008) sample includes two pairs of mid- to late-L dwarfs with very different near-IR 
colors. The authors find that the redder L dwarfs have less efficient sedimentation and therefore thicker 
cloud decks consisting of smaller particles, although gravity also may play a role for one pair. For one of 
the red L dwarfs Cushing et al. (2006) identified a broad absorption feature at 9--$11\,\mu$m which may be due 
to the presence of small silicate grains (Figure \ref{fig:mie}).

The difficulty in characterizing the L to T transition arises from our lack of understanding of the masses and effective temperatures of objects at various locations in the ultra-cool dwarf color-magnitude diagram.  It is not clear, for example, if the reddest field L dwarfs are more or less massive than bluer objects or if the
turn towards the blue in $J-K$ is mass dependent (although there are some indications that it may be (Metchev \& Hillenbrand 2006)).    There are two ways in which this shortcoming in current understanding could be addressed.  First,  observing the orbits of binary brown dwarfs  allows the total system mass to be measured.  Secondly, photometry leading to near-infrared color-magnitude diagrams for many clusters with a variety of ages and metalicities will constrain the nature of transition.  

To date, the cluster color magnitude diagram (CMD) has only reached the transition in the Pleiades and perhaps the Sigma Orionis clusters.  
Figure~\ref{fig:plds}b  shows the currently best available  CMD for the Pleiades.  Two objects in this figure can be seen to have turned towards the blue.  Deeper searches to fainter magnitudes in this cluster should soon reveal the expected downward turn to the fainter J  magnitudes apparent in the field CMD.  By constructing evolution models at the age of the Pleiades, it should be possible to constrain the mass at the turnoff from the red L sequence.  Given enough clusters of different ages, the turnoff effective temperature and gravity can be constrained, thus illuminating the dependence of the turnoff on gravity and perhaps metallicity.

JWST will be able to obtain moderate resolution ($R\sim 1000$) spectra on the current Pleiades candidates in Figure \ref{fig:plds} in about 2.5 hours at $S/N \sim 20$.  This spectral resolution should be sufficient to identify, for example, FeH and $\rm CH_4$ bands as they vary through the spectral sequence.  This combination of evolution models and spectra should tightly constrain the empirical cooling sequence.  Although this cluster is likely too large on the sky for efficient searching by JWST, NIRCAM could in principle find objects with masses as low as about $1\,\rm M_J$.  Surveys of more compact clusters would not reach to such low masses, but should nevertheless be deep enough to find many young T dwarfs that have already undergone the transition.

\subsection{The Latest T Dwarfs}
At this time only 16 very cool ($T_{\rm eff}< 900\,\rm K$) dwarfs with types T7 and later are known, and of these only four are T8 or later. New 
surveys that go fainter than 2MASS and SDSS have started or are planned, and several groups are attempting to 
push to later and cooler types (e.g. Warren et al. 2007). All but two of the very late T dwarfs are isolated (the 
exceptions are Gl 570 D, Burgasser et al. 2000, and HD 3651B, Mugrauer et al. 2007). Since age is unknown for field dwarfs and 
brown 
dwarf cool with time, observed spectra must be compared with models or spectra of fiducial objects, to constrain mass and age.
Since there are only a few T dwarfs with highly constrained properties, 
accurate model analysis is crucial for secure determination of gravity and 
hence mass at the bottom of the T sequence.   For field brown dwarfs with ages in the range $\sim$ 1-5~Gyr and masses of 20--50 
Jupiter-masses, effective temperature will lie in the range of $\sim$600--800~K.
To understand the physical parameters of  these 
elusive, cold, and low-mass dwarfs requires observation of their full spectral energy distribution.

\subsection{Vertical Mixing and Chemical Disequilibrium}
 It has long been understood that the abundances of molecules in Jupiter's atmosphere depart from the values predicted purely from equilibrium chemistry (Prinn \& Barshay 1977; Barshay \& Lewis 1978; Fegley \& Prinn 1985; Noll et al. 1988; Fegley \& Lodders 1994; Fegley \& Lodders 1996)\footnote{In stellar atmospheres, departures from thermochemical equilibrium can arise from interactions of atoms and molecules with the non-thermal radiation field (Hauschildt et al. 1997; Schweitzer, Hauschildt \& Baron 2000).  In brown dwarf atmospheres this effect is negligible.}.  Rapid upwelling can carry compounds from the deep atmosphere up into the observable regions of the atmosphere on time scales of hours to days.  When this convective timescale is shorter than the timescale for chemical reactions to reach equilibrium, then the atmospheric abundances will differ from those that would be found under pure equilibrium conditions.  The canonical example of this situation is carbon monoxide in Jupiter's atmosphere.  In Jupiter's cold and dense upper troposphere, carbon should almost entirely be found in the form of methane.  Deeper into the atmosphere, where temperatures and pressures are higher, CO should  be the principal carrier of carbon.  Rapid vertical mixing, combined with the strong C-O molecular bond, means that CO molecules can be transported to the observed atmosphere faster than chemical reactions can reduce the CO into $\rm CH_4$.  The observed enhancement of CO, combined with (uncertain) reaction rates places limits on the vigor of convective mixing in the atmosphere.

\paragraph{CO} Analyses of the 4.5 -- $5 \,\rm \mu m$  spectra of the T dwarfs Gl 229B and  Gl 570 D reveal an abundance of CO that is over 3 orders of magnitude larger than expected from chemical equilibrium calculations (Noll et al. 1997, Oppenheimer et al. 1998, Griffith \& Yelle 1999, Saumon et al. 2000), as anticipated by Fegley \& Lodders (1996).  Photometry in the $M$ band, which overlaps the CO band at $\rm 4.6\,\rm \mu m$, shows that an excess of CO may be a common feature of T dwarfs. Golimowski et al. (2004) have found that the $M$ band flux is lower than equilibrium models predict based on the $K$ and $L^\prime$ fluxes in all of the T dwarfs in their sample.  
The low $M$ band flux certainly arises from an excess of CO above that expected by equilibrium models.  Since CO is a strong absorber in the $M$ band, a brown dwarf can be much fainter in this band than would be predicted by equilibrium chemistry.  Equilibrium models predict that brown dwarfs and cool extrasolar giant planets should be bright at $M$ band, hence any flux decrement at this wavelength would have implications for surveys for cool dwarfs and giant planets. The degree to which this is a concern depends upon how the vigor of mixing declines with following effective temperature.  Hubeny \& Burrows (2007) recently have argued that this will not be a concern at temperatures below about 500 K, however, because the vigor of mixing falls with effective temperature.

\begin{figure}[argh4]
\centering
\includegraphics[scale=0.25,angle=0]{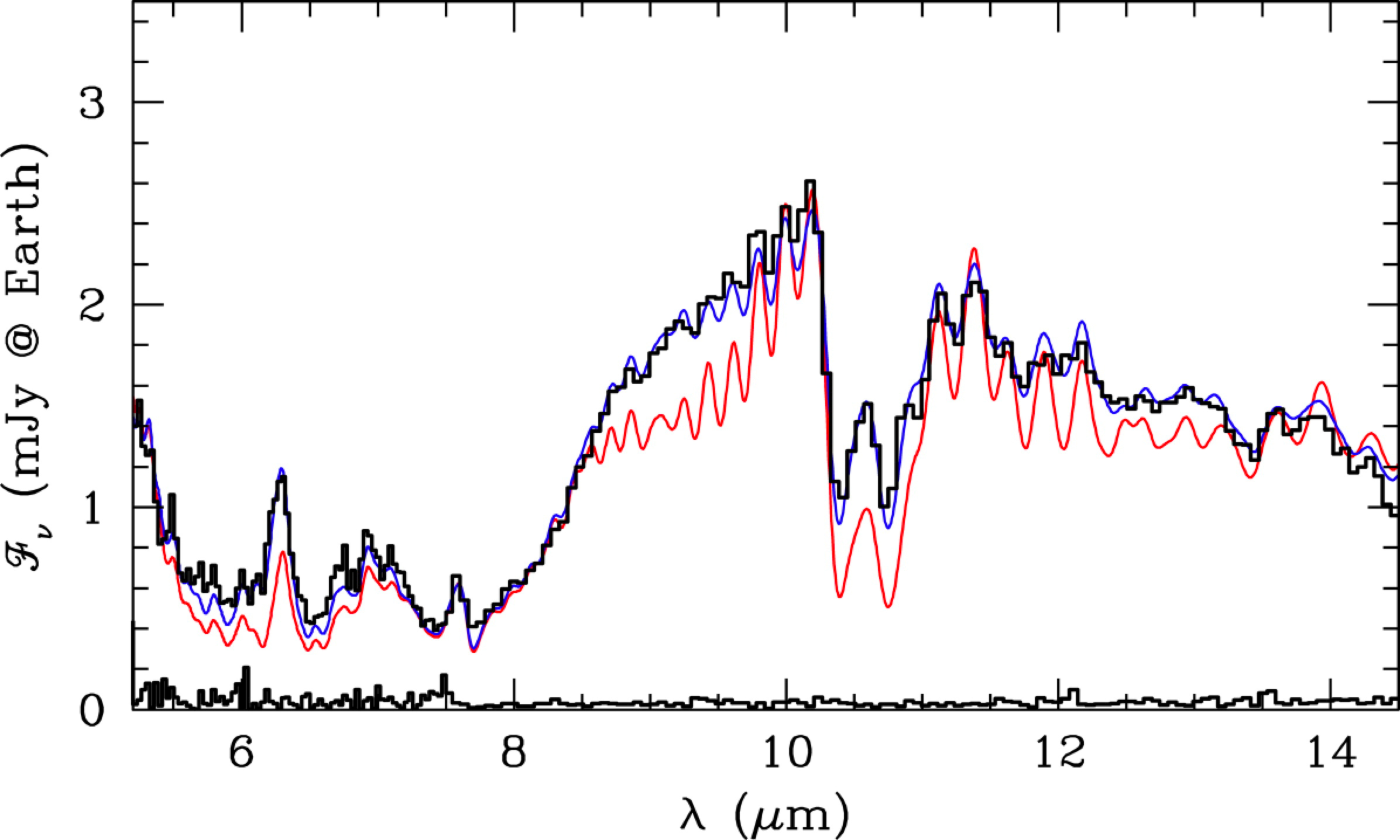}
\caption{Fits of the IRS spectrum of 2MASS J0415-0935  (Saumon et al. 2007) showing the difference between a model in chemical equilibrium and a model that includes vertical transport that drives the nitrogen and carbon chemistry out of equilibrium. The red thin curve is the best-fitting model in chemical equilibrium, and the blue thin curve is the best-fitting nonequilibrium model. The data and the noise spectrum are shown by the histograms (black). The uncertainty on the flux calibration of the IRS spectrum is $\pm 5\%$. The model fluxes, which have not been normalized to the data, are shown at the resolving power of the IRS spectrum.}
\label{fig:nh3}
\end{figure}

\paragraph{NH$_3$}Ammonia forms from $\rm N_2$ by the reaction $\rm N_2 + 3\,H_2 \Leftrightarrow 2\,NH_3$. $\rm N_2 + 3H_2$ is favored at low pressures and high temperatures because of higher entropy.  $\rm NH_3$ is favored at low temperatures, but since  molecular nitrogen is a strongly bound molecule, reactions involving this molecule typically have high reaction energies and proceed very slowly at low temperatures.  Like CO, $\rm N_2$ is favored at the higher temperatures found deep in brown dwarfs atmospheres.  Again, like CO, vigorous vertical transport can bring $\rm N_2$ in the upper atmosphere faster than it can be converted to $\rm NH_3$, resulting in an excess of $\rm N_2$ compared to the values expected from chemical equilibrium.  
Figure \ref{fig:nh3} illustrates the effect of mixing (Saumon et al. 2007) on the $10\,\rm\mu m$ ammonia band in the spectra of the T8 dwarf 2MASS0415-0935 (Burgasser et al. 2002a).
JWST will obtain higher resolution mid-infrared spectra than the {\em Spitzer} data analyzed by Saumon et al. (2007).  Higher spectral resolution on more targets will allow more in depth studies of atmospheric mixing.  Since the vertical profile of mixing also influences cloud particle sizes and optical depths (in L- and L  to T transition dwarfs), mapping out the eddy diffusion coefficient (which parameterizes mixing) as a function of mass and effective temperature will help to shed light on cloud dynamics as well as atmospheric chemistry.

\section{Opportunities for JWST}
As we have highlighted in the above sections, there are many unsolved problems in the study of ultracool dwarfs.  In this section we will briefly summarize some of the most promising avenues for JWST.

\subsection{Characterizing Rosetta Stone Dwarfs}
The characterization of most field brown dwarfs, particularly the L and early T dwarfs, is hampered by the dependency of model fits on the particulars of the cloud models used to generate model atmospheres and spectra for comparison to data.  Brown dwarfs of known mass, metallicity, and age are thus of particular importance as  calibrators for the entire brown dwarf cooling sequence.  This section discusses some opportunities for unearthing and deciphering `Rosetta Stone' ultracool dwarfs with known or easily deduced masses and effective temperatures. Such objects could turn the page to much greater understanding of our library of known L and T dwarfs.

\subsubsection{Color-magnitude diagram for clusters to low masses}
\begin{figure}[ht]
\centering
\includegraphics[scale=0.4,angle=0]{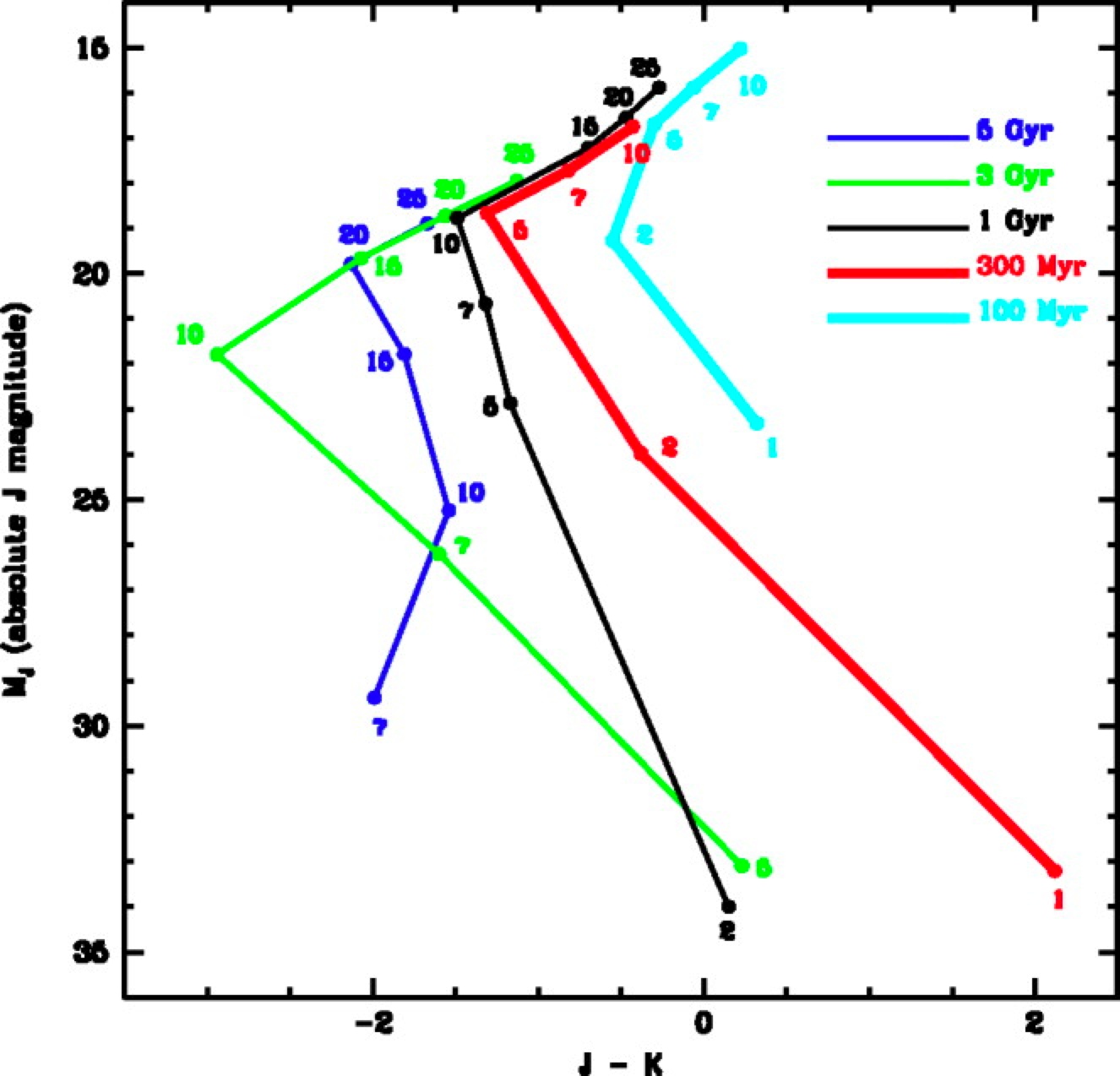}
\caption{Predicted absolute J magnitude ($M_J$) vs. $J-K$ color for a range of brown dwarf masses and ages.  The numbers by the symbols denote the masses of the objects in Jupiter mass units. In the Pleiades the JWST NIRCAM detection limit will be about 1 Jupiter mass. Figure and description from Burrows et al. (2003); see discussion therein for greater detail.}
\label{fig:y_dwarfs}
\end{figure}

 By providing cluster color-magnitude diagrams to very low masses (a Jupiter-mass or less) JWST will revolutionize our understanding of brown dwarf cooling in environments controlled for age and metallicity.  Comparison of spectra of objects with known properties to models will finally provide insight into the variation in cloud properties with mass and effective temperature. In the Pleiades, Casewell et al. (2007) have detected objects with masses as low as 11 Jupiter masses (Figure \ref{fig:plds}).  In each cluster a brown dwarf of a given mass will be found either earlier or later on its cooling track, depending on the cluster age.  Since the evolutionary cooling of brown dwarfs is well understood, spectra of cluster objects with known masses will  definitively connect spectral features with gravity for mid- to late L dwarfs.  In the Pleiades such a project should be straightforward for JWST as   NIRSPEC will be able to obtain $R\sim 1000$ JHK spectra of the known low-mass cluster members in as little as a few hours. 
A Jupiter-mass object will be about 6 magnitudes fainter in $J$ band.  Assuming JWST could survey a sufficiently large area to find candidates, it would define the brown dwarf cooling curve to a degree still not reached in the disk population.  Model photometric predictions (Burrows et al. 2003) for such objects are shown in Figure \ref{fig:y_dwarfs}.  The actual trajectory in color-magnitude space will ultimately depend upon the interplay of water clouds and atmospheric mixing with the emitted spectra.  Comparison of models such as those in the figure with data will test our understanding through this as-yet unexplored range of parameter space.
 
\subsubsection{Resolved Spectra of Close Binaries}
Binary stars have long served as astronomical workhorses, helping to reveal important details of stellar astrophysics.  Likewise binary brown dwarfs are also exceptionally useful.  The orbits of close L- and T-dwarf binaries allow the total system mass to be determined as Bouy et al. (2004) have done for the binary 2MASSW J0746425+2000321.  Assuming co-evality and equal metallicity combined with the total system mass and resolved spectra of the individual dwarfs furthermore allows such systems to elucidate the interpretation of brown dwarf spectra.  Many more tight binaries have been found by the combination of HST and ground based adaptive optics imaging (see summary in Bouy et al. 2008).  Orbital periods for many of these systems appear to be less than twenty years, so dynamical masses will be available during the JWST mission lifetime.  The combination of the ground-based astrometry and photometry and resolved NIRSPEC  high $R$ spectra of many of the individual objects, particularly for the tightest binaries, in these systems should provide important constraints  on models of brown dwarf evolution, atmospheric structure, and emergent spectra.

\subsubsection{Cloud Behavior from L to T to Y}
Perhaps the greatest single observational result that could drive improvements in understanding of the L to T transition would be a direct measurement of surface gravity (or almost equivalently, mass) and effective temperature of late L and early T dwarfs.  This would elucidate the dependence of the initiation of the L to T transition on mass and $T_{\rm eff}$.  Constraining the turnoff absolute magnitude in a variety of clusters of known ages and resolving the spectra of close binaries that have measurable dynamical masses would highly constrain the nature of the L to T transition.

Likewise as brown dwarfs further cool through the T sequence a number of open issues remain.  Although T dwarfs are generally modeled as being entirely cloud free, some models that include very thin cloud decks better reproduce the spectra and color of the T's.  As brown dwarfs cool through about 500 K, thin water clouds should appear high in their atmospheres.  With falling effective temperature these clouds are expected to thicken and begin to substantially alter the spectra of the Y dwarfs.  It is entirely possible that, like the departure of clouds in the late L dwarfs, the arrival of clouds in the early Y dwarfs will produce unexpected and perhaps rapid color and spectral changes.  Although we cannot yet identify what these changes might be, the same type of observations noted above will also be invaluable in constraining the Y dwarfs.  Since the optical and near-IR flux of the Y dwarfs is expected to rapidly decrease (Burrows et al. 2003), the signatures of the water clouds will be best obtained in the mid-IR by JWST.

\subsection{Spectra of Very Cool Objects}
\begin{figure}[ht]
\centering
\includegraphics[scale=0.42,angle=0]{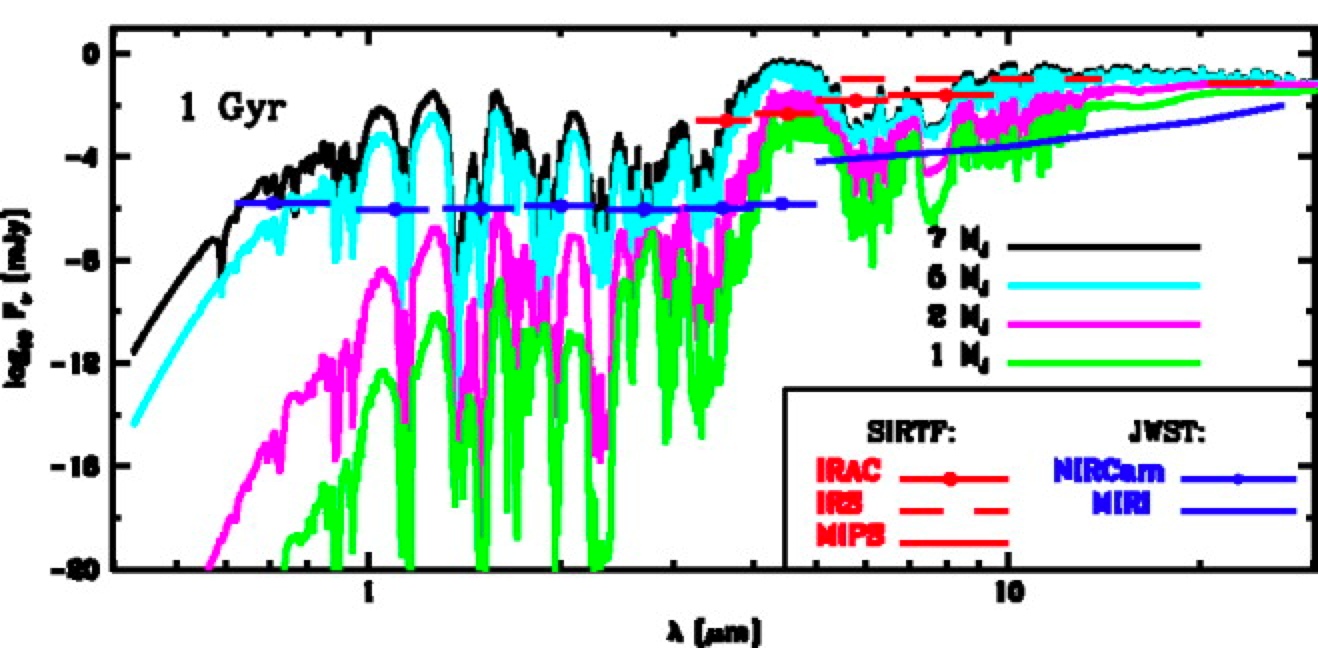}
\caption{Spectra (flux in millijanskys) vs. wavelength (in microns) for a range of brown dwarf masses at an age of 1 Gyr and a distance of 10 pc.  Superposed are the approximate point-source sensitivities for instruments on {\em Spitzer} (red) and {\em JWST} (blue). The JWST/NIRCam sensitivities are $5\,\sigma$ and assume an exposure time of $5\times10^4\,\rm sec$. The JWST/MIRI sensitivity curve from 5.0 to $27\,\rm \mu m$ is  $10\,\sigma$ and assumes an exposure time of $10^4\,\rm sec$. Figure and description from Burrows et al. (2003); see discussion therein for greater detail.}
\label{fig:cold}
\end{figure}

As of early 2008, the brown dwarf with the lowest estimated effective temperature is ULAS J0034-00 with $T_{\rm eff}\sim 650\,\rm K$ (Warren et al. 2007).  A number of ongoing and future searches will certainly find cool objects in the solar neighborhood (e.g., UKIDSS, Pan-STARRS, and the WISE mission).  In particular WISE will have sufficient sensitivity to detect a  $T_{\rm eff}\sim 200\,\rm K$ brown dwarf ($2\,\rm M_J$ at 1 Gyr) at a distance of about 2 pc.   Assuming such nearby targets are found, JWST will produce exquisite spectra that will be unobtainable from the ground.   A survey with NIRSPEC in high resolution mode with the YJH and LM gratings would nicely sample the spectra of cool field dwarfs.  For a 500K dwarf at 10 pc, we estimate an exposure of about a minute will provide a spectrum with S/N of about 100 in $M$ band.  A one hour exposure would be required for the same dwarf at 25 pc.   Detection limits and model spectra are shown  in Figure \ref{fig:cold}.

High quality spectra of cool dwarfs will be important for a number of reasons.  First, cold disk objects possess effective temperatures comparable to those of middle-aged to old extrasolar giant planets.  The disk population of brown dwarfs will thus provide ground truth for the spectral features that such cold objects exhibit (for example $\rm NH_3$ should appear in the near-IR (Saumon et al. 2001, Burrows et al. 2003, Leggett et al. 2008)).  These objects will also have water clouds.  The experience with the challenge of modeling silicate and iron clouds in L dwarfs alluded to above implies that water clouds will be no more tractable.  The cold disk population will thus provide a proving ground for exoplanet water cloud modeling, which is undoubtedly needed (see Marley et al. 2007 and references therein).

 Atmospheric mixing, long recognized in Jupiter's atmosphere, is also important for brown dwarfs (see \S4.5), yet the most diagnostic spectral region for this process (the CO band at $4.6\,\rm \mu m$) lies in a blind spot for {\em Spitzer} spectroscopy and most groundbased observatories, but not JWST. 
Thus the nearby disk brown dwarfs will elucidate the extent to which $M$ band flux of objects in this effective temperature range is impacted by excess atmospheric CO. 
Vertical mixing could be an important consideration for the direct detection of giant planets around nearby stars (Golimowski et al. 2004; Marley et al. 2006; Hinz et al. 2006). After the discovery of Gl 229B, Marley et al. (1996) suggested that a substantial 4 to $5\,\rm\mu m$ flux peak should be a universal feature of giant planets and brown dwarfs. This expectation, combined with a favorable planet/star flux ratio, has made the  band a favorite for planet detection (Burrows et al. 2005). However, groundbased and IRAC photometry suggests that  cool dwarfs are fainter in this region---and the $L$ band region is brighter---than predicted by equilibrium chemistry. Given these and other considerations, Leggett et al. (2007) suggested that the comparative advantage of ground-based searches for young, bright giant planets at $M$ band might, such as the searches planned with the JWST coronagraph, might be somewhat less than currently expected (see also Marley et al. 2007). 

Hubeny et al. (2007), however, recently predicted that the vigor of atmospheric mixing will  decline with effective temperature.  If this is indeed the case then $M$ band will remain a fruitful hunting ground for extrasolar giant planet coronagraphic imaging.  NIRCAM photometry of cold brown dwarfs will certainly illuminate this issue.

\subsection{Mid-IR Spectra Beyond {\em Spitzer}}

Examples of some of the best available {\em Spitzer} IRS mid-infrared spectra of L and T dwarfs are shown in Figures \ref{fig:mie} and \ref{fig:nh3}.  The spectral region between about 6 and $15\,\rm \mu m$ is important for a number of reasons.  First there are several strong molecular bands in this region, including water, methane, and ammonia.  As recounted above methane and ammonia are particularly sensitive to atmospheric mixing.  The Si-O vibrational band, seen in the opacity of small silicate grains (Figure \ref{fig:mie}), also may trace the arrival of silicate clouds.  Finally a number of other molecules, not yet detected in brown dwarf spectra, have absorption features in this range (Mainzer et al. 2007).

MIRI will produce much higher resolution and S/N spectra than the best data from IRS (see the sensitivity curve in Figure \ref{fig:cold}).  Higher quality spectra will allow for more robust detection of silicate features, perhaps including the sort of fine structure in the grain opacity seen in the lower panel of Figure \ref{fig:mie}, as well as for fine detail on the molecular features (see model prediction in Burrows et al. (2003)).  If silicates are indeed detected, high resolution spectra could in principle differentiate between the particular silicate species, including forsterite, enstatite, and even quartz ($\rm SiO_2$, Helling et al. 2006), and whether the grains are in crystalline or amorphous form (Figure \ref{fig:mie}).

\section{Conclusions}
The next decade holds the potential to substantially improve our fundamental understanding of ultracool dwarfs.  Ground and space based surveys for very cool dwarfs, including UKIDSS, Pan-STARRS, and the WISE mission as well as deep surveys of young clusters will provide a host of ultracool dwarf targets for JWST.  For cold nearby dwarfs JWST will provide unparalleled near- and especially mid-infrared spectra.  These observations will constrain the water clouds expected to be present in objects with $T_{\rm eff} < 500 \,\rm K$ and measure the degree of atmospheric mixing.  Both types of observations are highly relevant to the ultimate direct detection and characterization of extrasolar giant planets by coronagraphy.

In young clusters (e.g, the Pleiades and younger) JWST will provide exceptional quality spectra of many known cluster members and will have the capability of imaging objects down to about one Jupiter mass and below.  Such observations will constrain the evolutionary cooling tracks for dwarfs with lower gravities than most field objects and will tightly constrain the nature of the L to T transition by revealing its dependence on gravity.

Many other opportunities, including producing resolved spectra of tight binary dwarfs and searching for spectral signatures of condensates and low abundance gasses are also possible.  Combined with the inevitable unexpected discoveries, there is no doubt that JWST will bring brown dwarf astrophysics into the same highly constrained realm as stellar astrophysics.  Interpretting these expected datasets will undoubtedly require substantial improvements to atmosphere and evolution modeling, particularly cloud and chemical transport modeling of ultracool dwarf atmospheres.

\begin{acknowledgement}
The authors thank M.\ Cushing,  M.\ Liu, \& D.\ Saumon, for helpful conversations on the future of brown dwarf science and A.\ Burrows, Ch.\ Helling, X.\ Tielens and K.\ Zahnle for thoughtful comments on the manuscript.  We thank M.\ Cushing, M.\ Liu, and K.\ Lodders for preparing Figures 1, 6a \& 2, respectively and C.\ Nixon for kindly providing the Cassini CIRS spectrum of Jupiter for Figure 1.
\end{acknowledgement}

\section{References}

\noindent Ackerman, A.~S., \& 
Marley, M.~S.\ 2001, ApJ, 556, 872 

\noindent Barshay, S.~S., \& Lewis, J.~S.\ 1978, Icarus, 33, 593 

\noindent Bouy, H., et al.\ 2004, A\&A, 423, 341 

\noindent Bouy, H., et al.\ 2008, 
ArXiv e-prints, 801, arXiv:0801.4424 

\noindent Burgasser, A.~J., et 
al.\ 1999, ApJL, 522, L65 

\noindent  Burgasser, A.~J., et 
al.\ 2002a, ApJ, 564, 421

\noindent  Burgasser, A.~J., 
Marley, M.~S., Ackerman, A.~S., Saumon, D., Lodders, K., Dahn, C.~C., 
Harris, H.~C., \& Kirkpatrick, J.~D.\ 2002b, ApJL, 571, L151

\noindent Burgasser, A.~J., 
Geballe, T.~R., Leggett, S.~K., Kirkpatrick, J.~D., \& Golimowski, D.~A.\ 
2006, ApJ, 637, 1067 

\noindent Burrows, A., et al.\ 
1997, ApJ, 491, 856 

\noindent Burrows, A., Marley, 
M.~S., \& Sharp, C.~M.\ 2000, ApJ, 531, 438 

\noindent Burrows, A., Hubbard, 
W.~B., Lunine, J.~I., \& Liebert, J.\ 2001, Reviews of Modern Physics, 73, 
719 

\noindent Burrows, A., Sudarsky, 
D., \& Lunine, J.~I.\ 2003, ApJ, 596, 587

\noindent Burrows, A., Sudarsky, 
D., \& Hubeny, I.\ 2006, ApJ, 640, 1063 

\noindent Casewell, S.~L., 
Dobbie, P.~D., Hodgkin, S.~T., Moraux, E., Jameson, R.~F., Hambly, N.~C., 
Irwin, J., \& Lodieu, N.\ 2007, MNRAS, 378, 1131 

\noindent Cooper, C.~S., Sudarsky, 
D., Milsom, J.~A., Lunine, J.~I., \& Burrows, A.\ 2003, ApJ, 586, 1320 

\noindent Cushing, M.~C., et al.\  2006, ApJ, 648, 614 

\noindent Cushing, M.~C., et al.\ 
2008, ApJ, in press, arXiv:0711.0801 

\noindent Dahn, C.~C., et al.\ 2002, 
AJ, 124, 1170 

\noindent Fegley, B.~J., \& Lodders, K.\ 1994, Icarus, 110, 117 

\noindent Fegley, B., Jr., \& Prinn, R.~G.\ 1985, ApJ, 299, 1067 

\noindent Fegley, B.~J., \& Lodders, K.\ 1996, ApJL, 472, L37 

\noindent Freedman, R.~S., 
Marley, M.~S., \& Lodders, K.\ 2008, ApJSup, 174, 504

\noindent Geballe, T.~R., et al.\ 
2002, ApJ, 564, 466 

\noindent Golimowski, D.~A., 
et al.\ 2004, AJ, 127, 3516 

\noindent  Griffith, C.~A., \& Yelle, R.~V.\ 1999, ApJL, 519, L85 

\noindent Hanner, M.~S., Lynch, 
D.~K., \& Russell, R.~W.\ 1994, ApJ, 425, 274 

\noindent Helling, Ch., Thi, W.-F., Woitke, P., \& Fridlund, M.\ 2006, A\&A, 451, L9 

\noindent  Helling, Ch., et al.\ 2008a,  in prep. 

\noindent  Helling, Ch., Dehn, M., 
Woitke, P., \& Hauschildt, P.~H.\ 2008b, ApJL, 675, L105 

\noindent Hubeny, I., \& Burrows, A.\ 2007, ApJ, 669, 1248 

\noindent Kirkpatrick, J.~D., 
et al.\ 1999, ApJ, 519, 802 

\noindent Kirkpatrick, J.~D.\ 2005, 
Ann. Rev. Astron. \& Astrophys., 43, 195 

\noindent Kirkpatrick, J.~D.\ 2007, 
ArXiv e-prints, 704, arXiv:0704.1522 

\noindent Knapp, G.~R., et al.\ 
2004, AJ, 127, 3553

\noindent Kunde, V.~G., et al.\ 
2004, Science, 305, 1582 

\noindent Leggett, S.~K., et al.\ 
2000, ApJL, 536, L35

\noindent Leggett, S.~K., Saumon, 
D., Marley, M.~S., Geballe, T.~R., Golimowski, D.~A., Stephens, D., \& Fan, 
X.\ 2007, ApJ, 655, 1079

\noindent Liu, M.~C., Leggett, S.~K., 
Golimowski, D.~A., Chiu, K., Fan, X., Geballe, T.~R., Schneider, D.~P., 
\& Brinkmann, J.\ 2006, ApJ, 647, 1393 

\noindent  Lodders, K., \& Fegley, B.\ 2002, Icarus, 155, 393 

\noindent  Lodders, K., \& Fegley, B., Jr.\ 2006, Astrophysics Update 2, 1 

\noindent Lodieu, N., et al.\ 
2007, MNRAS, 379, 1423 

\noindent Mainzer, A.~K., et al.\ 
2007, ApJ, 662, 1245

\noindent Marley, M.~S., Saumon,  D., Guillot, T., Freedman, R.~S., Hubbard, W.~B., Burrows, A., \& Lunine, 
J.~I.\ 1996, Science, 272, 1919

\noindent Marley, M.\ 2000, From Giant 
Planets to Cool Stars, 212, 152 

\noindent Marley, M.~S., Fortney, 
J., Seager, S., \& Barman, T.\ 2007, Protostars and Planets V, 733 

\noindent Metchev, S.  \& Hillenbrand,
L. A. 2006, ApJ, 651, 1166

\noindent  Nakajima, T., 
Oppenheimer, B.~R., Kulkarni, S.~R., Golimowski, D.~A., Matthews, K., \& 
Durrance, S.~T.\ 1995, Nature, 378, 463

\noindent Noll, K.~S., Geballe, 
T.~R., \& Marley, M.~S.\ 1997, ApJL, 489, L87 

\noindent Noll, K.~S., Geballe, 
T.~R., Leggett, S.~K., \& Marley, M.~S.\ 2000, ApJL, 541, L75 

\noindent Oppenheimer, B.~R., 
Kulkarni, S.~R., Matthews, K., \& van Kerkwijk, M.~H.\ 1998, ApJ, 502, 932 

\noindent Patten, B.~M., et al.\ 
2006, ApJ, 651, 502 

\noindent Prinn, R.~G., \& Barshay, S.~S.\ 1977, Science, 198, 1031 

\noindent Roellig, T.~L., et al.\ 
2004, ApJS, 154, 418 

\noindent Saumon, D., Geballe, 
T.~R., Leggett, S.~K., Marley, M.~S., Freedman, R.~S., Lodders, K., Fegley, 
B., Jr., \& Sengupta, S.~K.\ 2000, ApJ, 541, 374 

\noindent  Saumon, D., Marley, 
M.~S., Cushing, M.~C., Leggett, S.~K., Roellig, T.~L., Lodders, K., \& 
Freedman, R.~S.\ 2006, ApJ, 647, 552 

\noindent Saumon, D., et al.\ 
2007, ApJ, 656, 1136 

\noindent Sharp, C.~M., \& Burrows, A.\ 2007, ApJSup, 168, 140 

\noindent Skrutskie, M.~F., et 
al.\ 2006, AJ, 131, 1163 

\noindent  Strauss, M.~A., et al.\ 
1999, ApJL, 522, L61 

\noindent Tinney, C.~G., 
Burgasser, A.~J., \& Kirkpatrick, J.~D.\ 2003, AJ, 126, 975 

\noindent Toppani, A., Libourel, 
G., Robert, F., Ghanbaja, J., 
\& Zimmermann, L.\ 2004, Lunar and Planetary Institute Conference Abstracts, 35, 1726 

\noindent Tsuji, T.\ 2002, ApJ, 575, 264 

\noindent Tsuji, T., \& Nakajima, T.\ 2003, ApJL, 585, L151 

\noindent  Tsuji, T., Nakajima, T., 
\& Yanagisawa, K.\ 2004, ApJ, 607, 511 

\noindent Tsuji, T.\ 2005, ApJ, 621, 1033 

\noindent Vrba, F.~J., et al.\ 2004, 
AJ, 127, 2948 

\noindent Warren, S.~J., et al.\ 
2007, MNRAS, 381, 1400 

\noindent Woitke, P., \& Helling, Ch.\ 2003, A \& A, 399, 297 

\noindent York, D.~G., et al.\ 2000, 
AJ, 120, 1579 

\end{document}